\documentclass[aps,preprint]{revtex4}%
\usepackage{amsfonts}
\usepackage{amsmath}
\usepackage{amssymb}
\usepackage{graphicx}%
\providecommand{\U}[1]{\protect\rule{.1in}{.1in}}
\begin{document}
\title[ ]{Thermodynamics of the Harmonic Oscillator: Derivation of the Planck Blackbody
Spectrum from Pure Thermodynamics }
\author{Timothy H. Boyer}
\affiliation{Department of Physics, City College of the City University of New York, New
York, New York 10031}

\begin{abstract}
In 1893, Wien applied the first two laws of thermodynamics to blackbody
radiation and derived his displacement theorem. \ Believing that the
information from thermodynamics had been exhausted, Planck turned to
statistical ideas in 1900 in order to provide a physical understanding for his
experimental--data-based interpolation giving the Planck spectrum without
zero-point radiation. \ Here we point out that the third law of thermodynamics
(which was developed in the early years of the 20th century) introduces
additional thermodynamic information regarding thermal radiation. \ The Planck
spectrum for thermal radiation can be derived from purely thermodynamic ideas
applied to the classical simple harmonic oscillator, since every radiation
mode takes a simple oscillator form. \ Thermodynamics alone implies the Planck
spectrum including zero-point energy without any need for quantum theory or
statistical ideas. \ It is pointed out that the inclusion of zero-point energy
involves more natural thermodynamics than its exclusion.

\end{abstract}
\maketitle

\section{Introduction}

At the end of the 19th century, there were only two laws of thermodynamics.
\ The first law introduced heat energy into the ideas of energy conservation.
\ The second law involved the existence of an entropy state function and an
absolute temperature scale. In 1893, these two laws were used by Wien to
derive his displacement theorem for the blackbody radiation
spectrum.\cite{Kuhn-W} However, Wien's displacement theorem did not solve the
blackbody radiation problem because there still remained an unknown function
of a single variable involving frequency divided by absolute temperature. \ In
1900, when Planck became convinced that the useful information from the laws
of thermodynamics had been exhausted, he turned to statistical mechanics to
give a physical basis for his interpolation guess for the blackbody radiation
spectrum.\cite{Hermann15} Planck's use of statistical ideas led to the
beginnings of quantum theory. \ It is striking that within a decade of
Planck's work, Nernst's heat theorem associated with the low-temperature
specific heats of solids gave the basis for what is now termed the third law
of thermodynamics.\cite{Nernst} Although the textbooks of thermodynamics and
statistical mechanics regard the third law as related to quantum
theory,\cite{Reif} the third law also fits into classical theory.\cite{B3rd}
\ Indeed it turns out that the use of all three laws of thermodynamics allows
a complete derivation of the Planck spectrum based upon pure thermodynamics
without any recourse whatsoever to a statistical theory. \ The basic
thermodynamics and mathematics of the analysis is elementary and can be
followed by an advanced undergraduate physics student.

\section{Outline of the Analysis}

Our analysis makes use of information about classical harmonic oscillators,
the laws of thermodynamics, and applications of analytic function theory. \ We
start by noting that a classical harmonic oscillator in a thermodynamic bath
can be regarded as a thermodynamic system with an energy, generalized force,
and temperature. \ The classical oscillator has an adiabatic invariant given
by the energy divided by the frequency, and this adiabatic invariant allows us
to recognize the generalized force associated with the work done by an
adiabatic change in the spring constant (or of the frequency). \ Thus a
classical harmonic oscillator in a heat bath can be treated as a very simple
thermodynamic system of two independent variables, energy $U$ and frequency
$\omega.$ \ Next we turn to the laws of thermodynamics, starting with the
existence of an empirical temperature and then using the first two laws to
establish that the entropy of the oscillator is a function of a single
variable involving the energy divided by the frequency. \ Next we connect the
behavior of the oscillator at large energy with the thermodynamic temperature.
\ The proportionality of the thermodynamic temperature and the oscillator
energy at high temperature allows us to obtain the functional form of the
oscillator entropy at large energy. \ We next assume that the entropy is an
analytic function of energy and apply the third law of thermodynamics to the
analytic form. \ We find that in order to satisfy the third law of
thermodynamics, the oscillator must have finite energy at zero thermodynamic
temperature. \ The behavior of the derivatives of the entropy near zero
temperature suggests singular terms in the entropy as a function of energy.
\ Making use of the singularities and comparing with the assumed
high-temperature form obtained earlier, we are able to obtain the Planck
spectrum with zero-point energy as the simplest entropy expression consistent
with the three laws of thermodynamics; any additional terms would involve
additional singularities in the complex plane. \ Finally we comment on the
classical thermodynamics which we have obtained in relation to the
corresponding quantum expressions. \ The article closes with brief historical
remarks regarding the third law of thermodynamics. \ 

\section{Harmonic Oscillators as Thermodynamic Systems}

Elementary discussions of thermodynamics often refer to gas systems with two
independent variables, pressure and volume, whereas the simple harmonic
oscillator is usually introduced only in connection with statistical
mechanics.\cite{therm} \ Nevertheless, the classical simple harmonic
oscillator provides a very simple thermodynamic system of two independent
variables, energy $U$ and frequency $\omega$. \ A mechanical oscillator has an
equation of motion $m\ddot{x}=-kx,$ an oscillator (angular) frequency
$\omega=(k/m)^{1/2},$ and an energy $U=(1/2)m\dot{x}^{2}+(1/2)kx^{2}.$ \ It
can be shown that the oscillator energy divided by the frequency,
$U/\omega=U(m/k)^{1/2},$ is an adiabatic invariant\cite{invar} for the
mechanical system, so that under a very slow change of the spring constant
$k$, the quantity $U(m/k)^{1/2}$ remains unchanged. \ Therefore we have%
\begin{equation}
0=dU\left(  m/k\right)  ^{1/2}-dk\left[  U\left(  m/k^{3}\right)
^{1/2}/2\right]  =dU/\omega-d\omega(U/\omega^{2}).\label{e1}%
\end{equation}
The expressions involving frequency $\omega$ are simpler in form than those
involving the spring constant $k$, and so we will refer to the change in the
oscillator frequency $\omega$\ rather than to the change in the spring
constant $k$. \ The change in energy $dU$ associated with the change $d\omega$
in the oscillator frequency $\omega$ must be provided by some external agent.
\ Then from Eq. (\ref{e1}), the work $dW$ done \textit{by the oscillator} is
just the negative of this agent's work, $dW=-$ $(U/\omega)d\omega,$ and
therefore\
\begin{equation}
X=-U/\omega\label{e2}%
\end{equation}
can be regarded as a generalized force associated with the oscillator. \ 

A classical harmonic oscillator can be regarded as coming to equilibrium in a
thermal bath and so providing a simple thermodynamic system. \ Equation
(\ref{e2}), connecting $X,~U,~$and $\omega,$ may be regarded as an equation of
state for the system. \ We will regard thermal \textit{radiation} as providing
the thermal bath, and each oscillator is treated as a small electric dipole
oscillator with a very small electric charge interacting with electromagnetic
radiation. \ At the end of the 19th century, Planck showed\cite{Lav} that a
small electric dipole oscillator came to equilibrium with random radiation
when the average energy of the oscillator matched the average energy of the
radiation-bath normal modes at the same frequency as that of the oscillator. \ 

\section{Zeroth Law: Empirical Temperature for a Harmonic-Oscillator System}

We will consider a collection of oscillators at a various mechanical
frequencies $\omega,$ where for each oscillator the average energy is given by
its energy $U,$ and the generalized force $X$ is given by $X=-U/\omega.$ \ The
zeroth law of thermodynamics introduces the idea of an empirical temperature
$\theta(U,\omega)$ associated with the average oscillator energy $U$ and the
mechanical frequency $\omega.$ \ If two oscillators of different frequencies
$\omega$ and $\omega^{\prime}$agree on the empirical temperature,
$\theta(U,\omega)=\theta(U^{\prime},\omega^{\prime}),$ then the two
oscillators will be in thermal equilibrium with the same thermal radiation
bath and in thermal equilibrium with each other. \ 

\section{First Law of Thermodynamics \ }

The first law of thermodynamics introduces the idea that heat energy $\delta
Q$ can be included in the conservation law for energy, as $\delta Q=dU+\delta
W$. \ For a harmonic oscillator, this conservation equation takes the form
\begin{equation}
\delta Q=dU-(U/\omega)\delta\omega\label{First}%
\end{equation}
where $\delta Q$ corresponds to the heat energy added to the oscillator, $U$
is the internal energy of the oscillator, and $-(U/\omega)\delta\omega$
corresponds to the work done by the oscillator. \ The internal energy
$U(\theta,\omega)$ of an oscillator is a function of the empirical temperature
$\theta$ and the frequency $\omega$ of the oscillator. \ We will now
\textit{define }the empirical temperature $\theta$ by choosing an oscillator
of a single preferred frequency $\omega_{\ast}$ and taking the average energy
$U_{\ast}$ of this oscillator as equal to the empirical temperature,
$\theta=U_{\ast}$. \ For this oscillator, as more heat energy $\delta Q$ is
added to the system, the energy $U_{\ast}$ will increase for fixed
$\omega_{\ast}$ according to Eq. (\ref{First}). \ In this fashion, we assure
that the empirical temperature $\theta=U_{\ast}=\theta(U_{\ast},\omega_{\ast
})=\theta(U,\omega)$ is an increasing function of energy for the oscillator at
the special frequency $\omega_{\ast}.$ \ 

\section{Second Law of Thermodynamics}

For our harmonic oscillator system, the second law of thermodynamics declares
that there exists an entropy state function $S(U,\omega)$ which is a function
of the average oscillator energy $U$ and of the oscillator frequency $\omega,$
such that%
\begin{equation}
TdS=dU-(U/\omega)d\omega\label{second}%
\end{equation}
where
\begin{equation}
T=(\partial U/\partial S)_{\omega} \label{Ttherm}%
\end{equation}
is a thermodynamic temperature which depends only upon the empirical
temperature $\theta.$ \ Here if we introduce the relation (\ref{Ttherm}) into
the second-law relation (\ref{second}) divided by $T$, then we find%
\begin{equation}
dS=\frac{1}{T}dU-\frac{U}{T\omega}d\omega=\left(  \frac{\partial S}{\partial
U}\right)  _{\omega}dU-\frac{U}{\omega}\left(  \frac{\partial S}{\partial
U}\right)  _{\omega}d\omega\label{dS}%
\end{equation}
\ However, if we regard $S$ as a function of $U$ and $\omega,$ so that
$dS=(\partial S/\partial U)_{\omega}dU+(\partial S/\partial\omega)_{U}%
d\omega,$ then the relation (\ref{dS}) requires%
\begin{equation}
\left(  \frac{\partial S}{\partial\omega}\right)  _{U}=-\frac{U}{\omega
}\left(  \frac{\partial S}{\partial U}\right)  _{\omega} \label{dSdw1}%
\end{equation}
which means that $S$ is a function of the single variable $(U/\omega).$ \ Thus
we have%
\begin{equation}
\left(  \frac{\partial S(U/\omega)}{\partial\omega}\right)  _{U}=-\frac
{U}{\omega^{2}}S^{\prime}(U/\omega)=-\frac{U}{\omega}\left(  \frac{\partial
S(U/\omega)}{\partial U}\right)  _{\omega}. \label{dSdw2}%
\end{equation}
where $S^{\prime}(U/\omega)$ refers to the derivative of $S(U/\omega)$ with
respect to its one argument $U/\omega.$ \ 

\section{Adiabatic Lines}

A reversible adiabatic process involving a harmonic oscillator involves work
being done without the addition of heat. \ Under such a process, the entropy
of the system remains unchanged since $\delta Q=TdS=0.$ \ Since for a
harmonic-oscillator thermodynamic system, the entropy $S$ is a function of
$U/\omega,$ the adiabatic lines will be straight lines through the origin in a
plot of average energy $U$ vs $\omega,$ or will be horizontal straight lines
in an indicator plot of the generalized force $U/\omega=-X$ vs $\omega.$

\section{Connecting Empirical and Thermodynamic Temperature}

In order to connect the empirical temperature $\theta$ with thermodynamic
temperature $T$, we can follow the basic form of analysis given in Sears and
Salinger for a gas system.\cite{SS} \ If we regard the average energy $U$ of
an oscillator as a function of frequency and thermodynamic temperature, then
we have $dU=(\partial U/\partial T)_{\omega}dT+(\partial U/\partial\omega
)_{T}d\omega.$ \ Substituting this expression into the second-law equation
(\ref{second}), we obtain
\begin{equation}
dS=\frac{1}{T}\left(  \frac{\partial U}{\partial T}\right)  _{\omega}%
dT+\frac{1}{T}\left[  \left(  \frac{\partial U}{\partial\omega}\right)
_{T}-\frac{U}{\omega}\right]  d\omega. \label{eq2}%
\end{equation}
Now comparing this expression with $dS=(\partial S/\partial T)_{\omega
}dT+(\partial S/\partial\omega)_{T}d\omega,$ gives%
\begin{equation}
\left(  \frac{\partial S}{\partial T}\right)  _{\omega}=\frac{1}{T}\left(
\frac{\partial U}{\partial T}\right)  _{\omega}\text{ \ \ and \ \ }\left(
\frac{\partial S}{\partial\omega}\right)  _{T}=\frac{1}{T}\left[  \left(
\frac{\partial U}{\partial\omega}\right)  _{T}-\frac{U}{\omega}\right]  .
\label{eq3}%
\end{equation}
Using the agreement of the mixed\ second-partial derivatives of $S$ with
respect to $T$ and $\omega$ in opposite orders, we have%
\begin{equation}
\left\{  \frac{\partial}{\partial\omega}\left[  \frac{1}{T}\left(
\frac{\partial U}{\partial T}\right)  _{\omega}\right]  \right\}
_{T}=\left\{  \frac{\partial}{\partial T}\left[  \frac{1}{T}\left\{  \left(
\frac{\partial U}{\partial\omega}\right)  _{T}-\frac{U}{\omega}\right\}
\right]  \right\}  _{\omega}. \label{eq4}%
\end{equation}
Next simplifying this expression and canceling the mixed second derivatives of
$U$ on the two sides, we have\cite{SS150}%
\begin{equation}
\left(  \frac{\partial U}{\partial\omega}\right)  _{T}=-\frac{T}{\omega
}\left(  \frac{\partial U}{\partial T}\right)  _{\omega}+\frac{U}{\omega},
\label{SS69}%
\end{equation}
or equivalently%
\begin{equation}
\frac{U}{\omega}=f(T/\omega) \label{WienD}%
\end{equation}
where $f(T/\omega)$ is some function of the single variable $T/\omega.$ \ The
information in Eq. (\ref{WienD}) corresponds to that of Wien's displacement
theorem.\cite{Bthermosc}

We now use equation (\ref{SS69}) to connect the empirical and thermodynamic
temperatures for the harmonic oscillator. \ The thermodynamic temperature $T$
is some function of the empirical temperature $\theta$ alone, so that
$(\partial\theta/\partial T)_{\omega}=d\theta/dT.$ \ Thus our equation
(\ref{SS69}) becomes%
\begin{equation}
\left(  \frac{\partial U}{\partial\omega}\right)  _{\theta}=-\frac{T}{\omega
}\left(  \frac{\partial U}{\partial\theta}\right)  _{\omega}\frac{d\theta}%
{dT}+\frac{U}{\omega}. \label{eq5}%
\end{equation}
This implies%
\begin{equation}
\frac{dT}{T}=\frac{1/\omega(\partial U/\partial\theta)_{\omega}}%
{U/\omega-(\partial U/\partial\omega)_{\theta}}d\theta=\frac{(\partial
U/\partial\theta)_{\omega}}{U-\omega(\partial U/\partial\omega)_{\theta}%
}d\theta. \label{dTT}%
\end{equation}
The left-hand side of this equation is a function of $T$ only, and therefore
the right-hand side must be a function of $\theta$ only and must be
independent of $\omega.$ \ Therefore we may choose to evaluate each term of
the right-hand side at $\omega=\omega_{\ast}$ where $\omega_{\ast}$ refers to
the special oscillator whose energy $U_{\ast}$ is used as the empirical
temperature $\theta.~$Thus when $\omega=\omega_{\ast},$ we have $\theta
=U_{\ast},$ $(\partial U_{\ast}/\partial\theta)_{\omega_{\ast}}=1,$ and
$d\theta=dU_{\ast}.$ \ The equation (\ref{dTT}) then becomes \
\begin{equation}
\frac{dT}{T}=\frac{1}{U_{\ast}-\omega_{\ast}(\partial U_{\ast}/\partial
\omega_{\ast})_{\theta}}dU_{\ast}. \label{d2}%
\end{equation}
Eventually for large enough energies $U_{\ast},$ we expect $U_{\ast}$ to
become much larger than the magnitude of $\omega_{\ast}(\partial U_{\ast
}/\partial\omega_{\ast})_{\theta}$ unless the slope of isothermals at
$\omega_{\ast}$ increases steadily with energy $U_{\ast},$ which seems
completely unphysical. \ \ \ Indeed, we expect that at high temperature, the
quantity $(\partial U_{\ast}/\partial\omega_{\ast})$ should become ever
smaller
\begin{equation}
\left(  \frac{\partial U_{\ast}}{\partial\omega_{\ast}}\right)  _{\theta
}\approxeq O\left(  \frac{\omega_{\ast}}{U_{\ast}}\right)  \text{ \ \ for
large }U_{\ast},
\end{equation}
corresponding to independence of the oscillator energy from the oscillator
frequency. \ But then we have
\begin{align}
\frac{dT}{T}  &  =\frac{1}{U_{\ast}[1-(\omega_{\ast}/U_{\ast})(\partial
U_{\ast}/\partial\omega_{\ast})_{\theta}]}dU_{\ast}\nonumber\\
&  =\left\{  \frac{1}{U_{\ast}}+O\left(  \frac{\omega_{\ast}^{2}}{U_{\ast}%
^{3}}\right)  \right\}  dU_{\ast}. \label{d3}%
\end{align}
Integrating gives
\begin{equation}
\ln T=\ln U_{\ast}+\ln C+O\left(  \frac{\omega_{\ast}^{2}}{U_{\ast}^{2}%
}\right)  \label{d4}%
\end{equation}
and
\begin{equation}
T=CU_{\ast}+O\left(  \frac{\omega_{\ast}^{2}}{U_{\ast}^{2}}\right)  \label{d5}%
\end{equation}
where $C$ is a constant. \ The scale constant $C$ connects thermodynamic
temperature $T$ and empirical temperature $\theta=U_{\ast}$ used above. \ The
usual choice corresponding to \textquotedblleft practical
units\textquotedblright\ is $C=1/k_{B}$ where $k_{B}$ is Boltzmann's constant.
\ However, in most of the analysis to follow, it is more convenient to choose
$C=1,$ corresponding to \textquotedblleft rational units,\textquotedblright%
\ where absolute temperature $T$ is measured in energy units and entropy $S$
is a dimensionless pure number.\cite{units} \ \ Since the choice of the
special frequency $\omega_{\ast}$ was arbitrary, we expect this relationship
to hold for large values of the oscillator energy, no matter what the
frequency of the oscillator \
\begin{equation}
T\rightarrow U\left[  1+\,O\left(  \frac{\omega^{2}}{U^{2}}\right)  \right]
\text{ for large }U. \label{TUO}%
\end{equation}
Thus at high temperatures, the energy of a harmonic oscillator becomes
independent of the frequency $\omega$ of the oscillator. \ This
high-temperature result agrees with the equipartition result found from
nonrelativistic classical statistical mechanics.

The entropy expression at all temperatures following from our Eq. (\ref{TUO})
follows from%
\begin{equation}
\left(  \frac{\partial S}{\partial(U/\omega)}\right)  _{\omega}=\frac{\omega
}{T}=\frac{\omega}{U\left[  1+O\left(  \omega^{2}/U^{2}\right)  \right]
}=\frac{\omega}{U}\left[  1+O\left(  \frac{\omega^{2}}{U^{2}}\right)  \right]
.
\end{equation}
Integrating once, we have
\begin{equation}
S(U/\omega)=\ln(U/\omega)+const+O\left(  \frac{\omega^{2}}{U^{2}}\right)  .
\label{Sgen}%
\end{equation}

\section{High-Temperature Entropy-Energy Connection Assumed at All
Temperatures}

The high-temperature limit $U\approx T$ in Eq. (\ref{TUO}) between oscillator
energy $U$ and thermodynamic temperature $T$ is actually consistent with the
first two laws of thermodynamics if the relation $T=U$ is assumed to hold for
all temperatures. \ Thus if we set
\begin{equation}
T=U\text{ \ \ \ assumed for all }T,\label{TeU}%
\end{equation}
then we have ($\partial U/\partial\omega)_{\theta}=0$ in Eq. (\ref{d2}) and
our analysis in Eqs. (\ref{d3})-(\ref{TUO}) goes through with no correction
term. \ We solve for the entropy function $S(U/\omega)$ using%
\begin{equation}
\frac{1}{T}=\left(  \frac{\partial S}{\partial U}\right)  _{\omega
}\label{temp}%
\end{equation}
which from Eq. (\ref{TeU}) becomes
\begin{equation}
\frac{\omega}{T}=\frac{\omega}{U}=\left(  \frac{\partial S}{\partial
(U/\omega)}\right)  _{\omega}\label{dSdU}%
\end{equation}
or
\begin{equation}
S(U/\omega)=\ln(U/\omega)+const\label{Shigh}%
\end{equation}
Thus at large values of energy, we expect the entropy to take the form in Eq.
(\ref{Shigh}), the first derivative to give%

\begin{equation}
0<\frac{1}{T}=\left(  \frac{\partial S}{\partial U}\right)  _{\omega}=\frac
{1}{U},\label{Thigh}%
\end{equation}
and the second derivative to give
\begin{equation}
0>\left(  \frac{\partial^{2}S}{\partial U^{2}}\right)  _{\omega}=\frac
{-1}{U^{2}}.\label{Chigh}%
\end{equation}
The function (\ref{Shigh}) indeed satisfies the criteria of the first two laws
of thermodynamics.

\section{Third Law of Thermodynamics and Low Temperatures}

Although the first two laws of thermodynamics hold generally at any
temperature, the third law of thermodynamics deals specifically with
thermodynamic behavior at low temperatures. \ In particular, there is no
assurance that the limiting expression $S(U/\omega)=\ln(U/\omega)+const.$ in
(\ref{Shigh}), which we have obtained for the entropy at high temperatures,
will continue to hold as the thermodynamic temperature tends toward absolute
zero. \ Indeed, the third law states that as $(\partial U/\partial S)_{\omega
}=T\rightarrow0,$ the values of the entropy fall toward zero, $S(U/\omega
)\rightarrow0.$ \ It is clear that the entropy relation in Eq. (\ref{Shigh})
does not satisfy the third law of thermodynamics. \ Thus equation
(\ref{Shigh}) leads to Eq. (\ref{Thigh}) and so gives $T=U$; yet taking
$U=T\rightarrow0_{+}$ in equation (\ref{Shigh}) gives entropy which diverges
to negative infinity, $S=\ln(U/\omega)+const\rightarrow-\infty$ as
$U\rightarrow0$ for fixed $\omega.$ \ Therefore in order to satisfy the third
law of thermodynamics, the entropy $S(U/\omega)$ of a harmonic oscillator must
depart from the high-temperature form in Eq. (\ref{Shigh}). \ We wish to
obtain the necessary modifications to the entropy function. \ 

\section{Zero-Point Energy}

The entropy function given by $S(U/\omega)=\ln(U/\omega)+const$ in Eq.
(\ref{Shigh}) is analytic on the positive real axis in $U/\omega,$ and the
corresponding temperature (involving \textit{derivatives} of the entropy)
\begin{equation}
\frac{\omega}{T}=\frac{dS}{d(U/\omega)}=\frac{\omega}{U}\label{temp2}%
\end{equation}
is \textit{positive} on the entire positive real axis for $U/\omega;$ however,
the entropy function itself is \textit{not positive} on the full real axis
since $\ln(U/\omega)\rightarrow-\infty$ as $U/\omega\rightarrow0$. \ Now we
expect the entropy $S(U/\omega)$ to be an analytic function of $U/\omega,$ so
that working from the high-temperature form, we expect the additional terms to
be given by a series expansion in inverse powers of $U/\omega$ so that
\begin{equation}
S(U/\omega)=\ln(U/\omega)+const+\sum_{n=1}^{\infty}\frac{a_{n}}{(U/\omega
)^{n+1}}\label{Sanal}%
\end{equation}
where the $a_{n}$ are constants and the sum involves powers to the $n+1$
because of the correction in Eq. (\ref{Sgen}) is $O(\omega^{2}/U^{2})$. \ As
$U/\omega$ becomes large, each of the terms $(U/\omega)^{-n-1}$ in the sum
goes to zero faster than $\ln(U/\omega)$ so that $\ln(U/\omega)$ indeed can
dominate at large values of $U/\omega.$ \ However, as $U/\omega$ decreases
toward zero, the terms $(U/\omega)^{-n-1}$ in the sum will diverge faster than
$\ln(U/\omega)$ and so will dominate the expression. \ Accordingly, the
function $S(U/\omega),$ which must be \textit{monotonically increasing} with
increasing $U/\omega$, cannot be \textit{positive and analytic} on the entire
positive real axis of $U/\omega.$ Instead, there must be some
\textit{positive} value of $U/\omega$ where $S(U/\omega)$ becomes zero. \ Thus
there must be a smallest \textit{positive} value of $U/\omega$ where the
thermodynamic temperature $T$ falls to zero. \ We will take this smallest
value as $U/\omega=\hbar/2.$ \ Thus every oscillator of frequency $\omega$ has
a smallest positive energy (a zero-point energy) given by $U_{0}%
=U(\omega,0)=\omega\hbar/2.$ Crucially, the third law of thermodynamics, when
applied to the harmonic oscillator system, immediately implies the idea of
zero-point energy at the\ absolute zero of temperature. \ Zero-point energy
for a harmonic oscillator is implicit in the third law of thermodynamics.

\section{Form of the Entropy Function}

The third law of thermodynamics requires that the entropy $S$\ approaches zero
as the temperature $T$ falls toward zero corresponding to $U/U_{0}$
approaching $1_{+}$. \ We will set $z=U/U_{0}=U/(\omega\hbar/2)$ so that the
entropy in Eq. (\ref{Sanal}) can be regarded as a function of $z$ \ with%
\begin{equation}
S(z)=\ln(z)+Const+\sum_{n=1}^{\infty}\frac{A_{n}}{z^{n+1}},\label{Sz}%
\end{equation}
where $Const$ and the coefficients $A_{n}$ are adjusted from $const$ and the
$a_{n}$ in Eq. (\ref{Sanal}) so as to absorb the constant $\hbar/2$ associated
with the oscillator's zero-point energy. \ Then it follows from Eq. (\ref{Sz})
that%
\begin{equation}
0<\frac{\omega\hbar/2}{T}=\frac{dS}{dz}=\frac{1}{z}+\sum_{n=1}^{\infty}%
\frac{-(n+1)A_{n}}{z^{n+2}},\label{Sz1}%
\end{equation}
and%
\begin{equation}
0>\frac{d^{2}S}{dz^{2}}=-\frac{1}{z^{2}}+\sum_{n=1}^{\infty}\frac
{(n+1)(n+2)A_{n}}{z^{n+3}},\label{Sz2}%
\end{equation}
where $S(z)\rightarrow0$ and $T(U,\omega)\rightarrow0$ as $z\rightarrow1_{+}.$
\ It follows from Eq.(\ref{Sz1}) that the entropy function must have a
divergent derivative $dS/dz$ as $z\rightarrow1$ since $(\omega\hbar
/2)/T\rightarrow\infty$ as $T\rightarrow0$ for fixed $\omega.$ Also, if the
first derivative of $S(z)$ has a singularity at $z=1,$ then the second
derivative must have an even stronger singularity. \ The weakest singularity
would be a logarithmic singularity, such as $dS/dz\sim\ln(z-1),$ but the next
derivative would give a first-order singularity as $d^{2}S/dz^{2}\sim1/(z-1)$.
\ And indeed for these singularities, the function $S$ itself could have a
finite limit at the singularity corresponding to $S\sim(z-1)\ln
(z-1)\rightarrow0$ for $z\rightarrow1_{+}.$ \ 

The known high-temperature behavior places further limits on the
low-temperature singular behavior. If we try a first-order singularity at
$z=1$ for the second derivative of $S(z),$ we must include an additional
first-order singularity at some other value $z=\alpha$ in order that the
second derivative of $S$ falls off at large $z$ as the second power of $z$ as
in Eq. (\ref{Sz2}). \ Thus we can write
\begin{align}
\frac{d^{2}S}{dz^{2}}  &  =-\frac{1}{(z-1)(z-\alpha)}+\sum_{n=1}^{\infty}%
\frac{(n+1)(n+2)b_{n}}{z^{n+3}}\nonumber\\
&  =\frac{-1}{1-\alpha}\left(  \frac{1}{z-1}-\frac{1}{z-\alpha}\right)
+\sum_{n=1}^{\infty}\frac{(n+1)(n+2)b_{n}}{z^{n+3}}, \label{Sz22}%
\end{align}
where the $b_{n}$ are constants, and so preserve the analytic form and
high-temperature limit in Eq. (\ref{Sz2}), provided that the value of $\alpha$
lies on the real axis with $\alpha<1,$ which is out of the physical region
$1<z,$ corresponding to oscillator energies above the zero-point energy. \ 

Integrating once in Eq. (\ref{Sz22}), we have%
\begin{align}
\frac{dS}{dz}  &  =\frac{-1}{1-\alpha}[\ln(z-1)-\ln(z-\alpha)]+\gamma
-\sum_{n=1}^{\infty}\frac{(n+1)b_{n}}{z^{n+2}}\nonumber\\
&  =\frac{-1}{1-\alpha}\left[  \ln z+\ln\left(  1-\frac{1}{z}\right)  -\ln
z-\ln\left(  1-\frac{\alpha}{z}\right)  \right]  +\gamma-\sum_{n=1}^{\infty
}\frac{(n+1)b_{n}}{z^{n+2}}\nonumber\\
&  =\frac{-1}{1-\alpha}\left[  -\frac{1}{z}+\frac{1}{2z^{2}}-...+\frac{\alpha
}{z}-\frac{\alpha^{2}}{2z^{2}}+...\right]  +\gamma-\sum_{n=1}^{\infty}%
\frac{(n+1)b_{n}}{z^{n+2}} \label{Sza}%
\end{align}
where $\gamma$ and the $b_{n}$ are constants. \ In order for this expression
(\ref{Sza})\ to go over to the high-temperature form given in Eq. (\ref{Sz1}),
we must have $\alpha=-1$ and $\gamma=0.$ \ Introducing these values into the
first line in Eq. (\ref{Sza}), we obtain
\begin{equation}
\frac{dS}{dz}=\frac{-1}{2}[\ln(z-1)-\ln(z+1)]-\sum_{n=1}^{\infty}%
\frac{(n+1)b_{n}}{z^{n+2}}. \label{Sz11}%
\end{equation}
Now integrating equation (\ref{Sz11}) gives
\begin{align}
S(z)  &  =\frac{-1}{2}[(z-1)\ln(z-1)-(z+1)\ln(z+1)]+const^{\prime}+\sum
_{n=1}^{\infty}\frac{b_{n}}{z^{n+1}}\nonumber\\
&  =\frac{-1}{2}\left\{  (z-1)\ln z+(z-1)\ln\left(  1-\frac{1}{z}\right)
-(z+1)\ln z-(z+1)\ln\left(  1+\frac{1}{z}\right)  \right\} \nonumber\\
&  +const^{\prime}+\sum_{n=1}^{\infty}\frac{b_{n}}{z^{n+1}}\nonumber\\
&  =\ln z+\frac{-1}{2}\left\{  (z-1)\left(  \frac{-1}{z}+\frac{1}{2z^{2}%
}-...\right)  -(z+1)\left(  \frac{1}{z}-\frac{1}{2z^{2}}+...\right)  \right\}
+const^{\prime}+\sum_{n=1}^{\infty}\frac{b_{n}}{z^{n+1}}\nonumber\\
&  =\ln z+const^{\prime\prime}+\sum_{n=1}^{\infty}\frac{c_{n}}{z^{n+1}}
\label{Szac}%
\end{align}
where $const^{\prime},$ $const^{\prime\prime},$ and $c_{n}$ are all
constants.~\ The final line in Eq. (\ref{Szac}) has a large-$z$ analytic
behavior consistent with that in Eq. (\ref{Sz}). \ 

In our analysis starting with the third law of thermodynamics, we see the need
for certain singular terms to appear in the derivatives of $S(z),$ and we can
then use integrals to trace back the implications for their appearance in the
function $S(z)$ itself. \ If we take only the terms required by the
singularities at low temperature while still maintaining the high-temperature
limit, then we have the expression for the entropy
\begin{equation}
S\left(  z\right)  =\frac{-1}{2}[(z-1)\ln(z-1)-(z+1)\ln(z+1)]-\ln2 \label{Szz}%
\end{equation}
which vanishes as $z$ approaches $1,$ $z\rightarrow1_{+}.$ \ This function
(\ref{Szz}) can be rewritten in the high-temperature form given in Eq.
(\ref{Sz}) as%

\begin{align}
S(z)  &  =\frac{-1}{2}[(z-1)\ln(z-1)-(z+1)\ln(z+1)]-\ln2\nonumber\\
&  =\frac{-1}{2}\left\{  (z-1)\ln z+(z-1)\ln\left(  1-\frac{1}{z}\right)
-(z+1)\ln z-(z+1)\ln\left(  1+\frac{1}{z}\right)  \right\}  -\ln2\nonumber\\
&  =\ln z+\frac{-1}{2}\left\{  (z-1)\left(  \frac{-1}{z}+\frac{1}{2z^{2}%
}-...\right)  -(z+1)\left(  \frac{1}{z}-\frac{1}{2z^{2}}+...\right)  \right\}
-\ln2\nonumber\\
&  =\ln z+1-\ln2-\frac{1}{6z^{2}}-\frac{1}{20z^{4}}-\frac{1}{42z^{6}%
}-...\text{ \ for \ }1<z. \label{Szza}%
\end{align}
The derivatives follow as%

\begin{align}
\frac{dS}{dz}  &  =\frac{-1}{2}[\ln(z-1)-\ln(z+1)]=\frac{1}{2}\ln\left(
\frac{z+1}{z-1}\right) \nonumber\\
&  =\frac{1}{z}+\frac{1}{3z^{3}}+\frac{1}{5z^{5}}+...>0\text{ \ for }1<z
\label{Szzb}%
\end{align}
and%
\begin{align}
\frac{d^{2}S}{dz^{2}}  &  =\frac{-1}{2}\left(  \frac{1}{z-1}-\frac{1}%
{z+1}\right)  =\frac{-1}{z^{2}-1}\nonumber\\
&  =-\frac{1}{z^{2}}-\frac{1}{z^{4}}-\frac{1}{z^{6}}-...<0\text{ \ for }1<z.
\label{Szzc}%
\end{align}
Indeed by repeated differentiation, we can show that $S(z)$ and all its
derivatives are monotonic functions of $z$ for $1<z$ along the real line.
\ When considered in the complex $z$-plane, $S(z)$ in Eq. (\ref{Szz}) has
singularities at $z=1$ and $z=-1$ and at no other points in the complex plane.
Adding any further analytic functions to the expression for $S(z)$ in Eq.
(\ref{Szz}) would introduce additional singularities in the complex plane.
\ Thus the solution obtained in Eq. (\ref{Szz}) is the simplest solution which
meets the requirements of the three laws of thermodynamics.

\section{Obtaining the Planck Spectrum}

We have obtained the expression for the entropy $S(z)=S[U/(\omega\hbar/2)]$ of
a harmonic oscillator in terms of the single variable involving oscillator
energy divided by its zero-point energy. \ However, it is usual to give the
spectrum for blackbody radiation in terms thermodynamic temperature in
practical units $k_{B}T$ and frequency $\omega$ with the spectral density
\begin{equation}
\rho(\omega,T)=(\pi^{2}\omega^{2}/c^{3})U(\omega,T)\label{rho}%
\end{equation}
where $U(\omega,T)$ is the average energy of a harmonic oscillator (or
radiation mode) given in terms of frequency $\omega$ and thermodynamic
temperature $T$. \ We obtain this form by using Eq. (\ref{Szzb}) rewritten in
practical units as%
\begin{equation}
\frac{\omega\hbar/2}{k_{B}T}=\frac{d[S/k_{B}]}{d[U/(\omega\hbar/2)]}=\frac
{1}{2}\ln\left(  \frac{U/(\omega\hbar/2)+1}{U/(\omega\hbar/2)-1}\right)
.\label{rho1}%
\end{equation}
Taking the exponential, we find%
\begin{equation}
\exp\left(  \frac{\hbar\omega}{k_{B}T}\right)  =\left(  \frac{U/(\omega
\hbar/2)+1}{U/(\omega\hbar/2)-1}\right)  .\label{rho2}%
\end{equation}
Now solving for $U/(\omega\hbar/2),$ we obtain%
\begin{equation}
U/(\omega\hbar/2)=\frac{\exp[\hbar\omega/(k_{B}T)]+1}{\exp[\hbar\omega
/(k_{B}T)]-1}=\coth\left(  \frac{\hbar\omega}{2k_{B}T}\right)  \label{rho3}%
\end{equation}
or%

\begin{equation}
U=\frac{1}{2}\hbar\omega\coth\left(  \frac{\hbar\omega}{2k_{B}T}\right)
=\frac{\hbar\omega}{\exp[\hbar\omega/(k_{B}T)]-1}+\frac{1}{2}\hbar\omega,
\label{Planck}%
\end{equation}
which is the familiar Planck form, including zero-point energy $U(\omega
,0)=(\hbar/2)\omega.$ \ Indeed by the use of purely thermodynamic ideas, we
are led to the Planck formula including zero-point radiation for blackbody radiation.

\section{Regarding the Absence of Zero-Point Energy in the Interpolation
Formula Obtained by Planck}

In classical physics, a dipole harmonic oscillator must come to equilibrium
with the surrounding radiation bath where the average energy of the oscillator
matches the average energy of the bath radiation modes at the same frequency
as the oscillator frequency. \ Thus the harmonic oscillator treated in the
present article must be in equilibrium with a thermal bath which includes both
zero-point radiation and thermal radiation above the zero-point radiation as
given in Eq. (\ref{Planck}). \ \ 

The unambiguous situation for classical physics is in contrast to the
confusion which sometimes occurs in quantum physics. \ The extrapolation
expression based upon experimental data obtained by Planck in 1900 was%

\begin{equation}
U_{P}=\frac{\hbar\omega}{\exp[\hbar\omega/(k_{B}T)]-1}, \label{PlanckN}%
\end{equation}
and included no zero-point energy. \ This result without zero-point energy is
often repeated in the textbooks\cite{qm} as the energy of the radiation modes
of blackbody radiation. \ Indeed at high temperature, the Planck form
(\ref{PlanckN}) goes over to
\begin{equation}
U_{P}=k_{B}T-\frac{\hbar\omega}{2}+O(\frac{\hbar\omega}{k_{B}T}) \label{PHT}%
\end{equation}
which still subtracts the zero-point energy compared with the equipartition
result $U=k_{B}T$. \ This situation is in contrast to the analysis in the
present article where the oscillator energy at high temperatures was assumed
in Eq. (\ref{TUO}) to go over to the equipartion result $U=k_{B}%
T+O(\hbar\omega/k_{B}T).$ Modern quantum mechanics does include a zero-point
energy for a quantum harmonic oscillator.\cite{G46} \ However, the quantum
oscillator does not radiate when it is in its energy eigenstates, and so the
association with zero-point energy in the radiation field is not obvious. \ 

One must recall that Planck worked from thermodynamics insofar as requiring
the Wien-displacement-theorem form, which depended upon the first two laws of
thermodynamics. \ His blackbody energy expression (\ref{PlanckN}) was obtained
as an interpolation (involving experimental data) between the Wien radiation
formula, which seemed to fit the high-frequency data well, and the new
experimental results of Rubens and Kurlbaum, which suggested a different
low-frequency behavior.\cite{H12} \ Once he had obtained his interpolation
function in Eq. (\ref{PlanckN}), which provided an excellent fit to the
experimental data, he knew that he had to have an entropy expression of the
form\cite{H12}%
\begin{equation}
S(U_{p},\omega)=k_{B}\left[  \left(  \frac{U_{P}}{\hbar\omega}+1\right)
\ln\left(  \frac{U_{P}}{\hbar\omega}+1\right)  -\left(  \frac{U_{P}}%
{\hbar\omega}\right)  \ln\left(  \frac{U_{P}}{\hbar\omega}\right)  \right]
\label{SN}%
\end{equation}
where $U_{P}$ was the thermal\ energy of Planck's oscillator in equilibrium
with the radiation spectrum measured by the experimentalists. This equation
(\ref{SN}) is the same as our Eq. (\ref{Szz}) when we introduce the missing
constant $k_{B}$ so as to give
\begin{equation}
S[U/(\omega\hbar/2)]=k_{B}\left[  \frac{1}{2}\left(  \frac{U}{\hbar\omega
/2}+1\right)  \ln\left(  \frac{U}{\hbar\omega/2}+1\right)  -\frac{1}{2}\left(
\frac{U}{\hbar\omega/2}-1\right)  \ln\left(  \frac{U}{\hbar\omega/2}-1\right)
-\ln2\right]  , \label{SUw}%
\end{equation}
and we note the connection between the thermodynamic oscillator energy $U$ and
the thermal energy $U_{P}$\ used by Planck,%
\begin{equation}
U=U_{P}+\hbar\omega/2. \label{UUP}%
\end{equation}
Planck's energy $U_{P}$ is the thermal energy above the zero-point energy.
\ The Planck form in Eq. (\ref{PlanckN}) represents a shift in the oscillator
energy over to $U_{P}$ where the entropy $S[U_{P}/(\omega\hbar/2)]$ goes to
zero as the energy $U_{P}$ goes to zero.

Equations (\ref{SN}) and (\ref{SUw}) give the oscillator entropy $S(U,\omega)$
as a function of the energy divided by the frequency. \ However, we can also
obtain the entropy $S(\omega,T)$ as a function of frequency $\omega$ and
thermodynamic temperature $T$ by noting from Eq. (\ref{rho2}) that $\omega/T$
is a function of $U/\omega$ so that \ Eq. (\ref{rho1}) and the derivative of
Eq. (\ref{rho3}) can be combined as%
\begin{equation}
\frac{dS(\omega/T)}{d(\omega/T)}=\left(  \frac{dS(U/\omega)}{d(U/\omega
)}\right)  \left[  \frac{d(U/\omega)}{d(\omega/T)}\right]  =\left(
\frac{\hbar\omega}{2T}\right)  \left[  -\frac{\hbar}{2k_{B}}\text{csch}%
^{2}\left(  \frac{\hbar\omega}{2k_{B}T}\right)  \right]  \text{ }. \label{F1}%
\end{equation}
Then integrating gives%
\begin{equation}
S(\omega/T)=k_{B}\left\{  -\ln\left[  \sinh\left(  \frac{\hbar}{2}\frac
{\omega}{k_{B}T}\right)  \right]  +\frac{\hbar}{2}\frac{\omega}{k_{B}T}%
\coth\left(  \frac{\hbar}{2}\frac{\omega}{k_{B}T}\right)  -\ln2\right\}
\label{F2}%
\end{equation}
where the constant of integration is chosen so that $S\rightarrow0$ as
$T\rightarrow0.~\ $The zero-point energy makes no contribution to the entropy,
and therefore both the Planck form (\ref{PlanckN}) without zero-point energy
and the thermodynamic form (\ref{Planck}) including zero-point energy give the
same entropy dependence upon frequency and temperature for the harmonic
oscillator system.\ 

The discrepancy between the forms in Eqs. (\ref{Planck})\ and (\ref{PlanckN})
reappears in the calculation of the generalized force associated with the
harmonic oscillator system. \ The generalized force $X$ for the classical
simple harmonic oscillator system appears in the fundamental thermodynamic
relation $\delta Q=TdS=dU+Xd\omega.$ \ From the adiabatic invariant of the
mechanical system, we saw that $X=-U/\omega$ where $U$ is the mechanical
energy of the mechanical oscillator. \ In thermodynamics, we expect (from
$TdS=dU+Xd\omega)$ that the generalized force is given by
\begin{equation}
X=-\left(  \frac{\partial U}{\partial\omega}\right)  _{S}. \label{F3}%
\end{equation}
Now from equation (\ref{SUw}), we see that if $S$ is held constant, then
$U/\omega$ must be constant,\ so $U=\beta\omega$ where $\beta~$is a constant.
\ But then the generalized force follows from $(\partial U/\partial\omega
)_{S}=\beta=U/\omega,$ so that%

\begin{equation}
X=-\left(  \frac{\partial U}{\partial\omega}\right)  _{S}=-\frac{U}{\omega
}=-\frac{U_{P}+\hbar\omega/2}{\omega}=-\frac{U_{P}}{\omega}-\frac{\hbar}{2},
\label{F4}%
\end{equation}
involving the same connection $X=-U/\omega$ as was given in Eq. (\ref{e2}).
\ The contribution of the classical zero-point energy is absent in the Planck
form (\ref{PlanckN}). \ Thus the use of the Planck energy $U_{P},$ which
excludes the zero-point energy, requires the introduction of an additional
constant $-\hbar/2$ at all temperatures in the expression (\ref{F4}) for the
generalized force $X$. \ Even at low temperature $T\approx0,$ where the Planck
thermal energy $U_{P}$ approaches zero, there is still a non-zero generalized
force given by $X\approx-\hbar/2.$ \ In classical physics, this force arises
from the random radiation associated with zero-point radiation. \ In quantum
theory, this generalized force is a "quantum force" which is different from
that associated with thermal energy. \ As far as classical physics is
concerned, the omission of the zero-point energy in the expression
(\ref{PlanckN}) is misleading in suggesting that thermodynamic effects vanish
as the temperature goes toward absolute zero. \ The inclusion of an explicit
zero-point energy in the system energy, as in Eq. (\ref{Planck}), fits with
the thermodynamics of classical physics and is far more natural than its
absence. \ 

It is also interesting that the second derivative of the entropy with respect
to the energy in Eq. (\ref{Szzc}) can be rewritten in the form associated with
Einstein's ideas of energy fluctuations.\cite{Einstein}\cite{B69} \ Thus Eq.
(\ref{Szzc}) becomes in practical units%
\begin{equation}
\frac{d^{2}S}{dU^{2}}=-\frac{k_{B}}{U^{2}-U_{0}^{2}}=-\frac{k_{B}}%
{U^{2}-(\hbar\omega/2)^{2}}%
\end{equation}
and associates the entropy $S$ with the energy fluctuations (of the oscillator
or radiation mode) above the zero-point energy fluctuations. \ However, this
equation can also be rewritten in terms of the energy $U_{P}$ corresponding to
Planck's shift of the energy by the amount of the zero-point energy,
$U=U_{P}+U_{0}=U_{P}+\hbar\omega/2.$ \ In the form using the energy $U_{P}$,
the equation becomes%
\begin{equation}
\frac{d^{2}S}{dU^{2}}=-\frac{k_{B}}{(U_{P}+U_{0})^{2}-U_{0}^{2}}=-\frac{k_{B}%
}{U_{P}^{2}+2U_{0}U_{P}}=-\frac{k_{B}}{U_{P}^{2}+\hbar\omega U_{P}}.
\end{equation}
It is this last form involving $U_{P}$ which corresponds to the analysis given
by Einstein in 1909. \ Einstein interpreted the fluctuations as involving a
"wave-like" contribution $U_{P}^{2}$ and a "particle-like" (photon)
contribution $\hbar\omega U_{P}$. \ Connecting the energy $U_{P}$ to the
number $N$\ of photons of frequency $\omega$ as $U_{P}=N\hbar\omega$, the
entropy (which follows as in Eq. (\ref{SN})) goes to zero as the number $N$ of
photons of energy $\hbar\omega$ goes to zero. \ 

Planck and Einstein's use of the thermal energy $U_{P}$ without zero-point
energy was natural at the turn of the 20th century since the experimental
bolometric measurements of thermal radiation involved only the contributions
of sources at temperatures above the temperature of the detectors, and thus
did not measure zero-point energy. \ A century later, by
Casimir-force\cite{Casimir} measurements,\cite{exp} it has become possible
(within a classical understanding) to measure \textit{all} the radiation
(including the zero-point radiation) surrounding two parallel conducting
plates. \ Although the generalized force $X$ given in Eq. (\ref{F4}) for a
hypothetical simple harmonic oscillator (associated with the energy dependence
on the frequency $\omega$) may not be accessible at the atomic level, the
generalized force involving random radiation surrounding two conducting
parallel plates is indeed measurable.\cite{exp} \ \ The forces between
conducting parallel plates are the radiation-mode analogues of the oscillator
generalized force $X.$ \ The Planck form (\ref{Planck}) involving photons for
the blackbody radiation spectrum, which appears in the modern physics
textbooks and in quantum statistical mechanics,\cite{qm} does not include
zero-point energy and goes to zero as the thermodynamic temperature goes to
zero. \ Thus the photons of the Planck spectrum make no contribution to the
Casimir forces between conducting parallel plates at zero temperature.
\ Quantum theory must introduce a separate "quantum mechanical" calculation
for the Casimir force at zero temperature, and then adds finite-temperature
corrections involving thermal excitations (photons) at non-zero temperature.
This is exactly analogous to the need for a "quantum mechanical" generalized
force $X$ for an oscillator at zero temperature. Classical theory, which uses
the full thermodynamic expression (\ref{Planck}) including the zero-point
radiation for the energy of a radiation mode, needs only a single calculation
to obtain the Casimir force at any temperature.\cite{Lam} \ In addition to the
need to separate a "quantum behavior" at zero temperature and a thermal
contribution (due to photons) at positive temperature, there are also
arguments among quantum theorists as to whether or not the Casimir effect at
zero temperature should actually be associated with quantum zero-point
energy.\cite{Jaffe} \ Within classical physics, there is no ambiguity; Casimir
forces are associated with changes in the energy of zero-point
radiation.\cite{B72}

\section{Discussion}

For over a century now, physicists have presented various theoretical
explanations for the Planck spectrum, some based upon quantum
theory\cite{quantum} and some based upon classical theory.\cite{classical}
\ However, one may wonder what is the minimal set of assumptions necessary to
derive the experimentally observed spectrum. \ Thermodynamics , which was
developed during the 19th century, has provided fundamental ideas of absolute
generality. \ The first two laws of thermodynamics provided the basis for
Boltzmann's derivation of the Stefan-Boltzmann relation $\mathcal{U=}%
a_{S}T^{4}V$ connecting the thermal energy $\mathcal{U}$ with the temperature
$T$ and volume $V$ for the thermal radiation in a closed container at thermal
equilibrium. \ These same two laws of thermodynamics, applied to the
reversible adiabatic compression of thermal radiation, allowed Wien to derive
his displacement theorem in 1893. \ Planck's enthusiasm for thermodynamics led
him to consider the problem of blackbody radiation at the end of the 19th
century; however, he became convinced that the information available from the
first two laws of thermodynamics had been exhausted by Wien in obtaining the
displacement theorem. \ Therefore he turned to statistical ideas after his
successful experimental-data-based interpolation giving the Planck spectrum
(without zero-point radiation). \ When the third law of thermodynamics was
introduced in the first decade of the 20th century, the law was associated
with the developing quantum theory. Apparently physicists never looked back to
see whether thermodynamics (now including the third law) might provide the
crucial missing function in Wien's displacement theorem.

The thermodynamics of thermal radiation is tightly connected with the
thermodynamics of the harmonic oscillator since the behavior of each normal
mode of oscillation for radiation takes the simple-harmonic-oscillator form.
\ Indeed, it was Planck who first pointed out that a small classical dipole
oscillator would come to thermal equilibrium with an average energy which was
the same as the average energy of the random-radiation normal modes at the
same frequency as the oscillator. \ Now the thermodynamics of the harmonic
oscillator is virtually never discussed in connection with fundamental
thermodynamics rather than in connection with statistical mechanics. \ Yet the
thermodynamics of the simple harmonic oscillator takes a very simple form.
\ \ Fifteen years ago, it was pointed out\cite{Bthermosc} that the application
of the first two laws of thermodynamics to a harmonic oscillator could lead
easily to the information corresponding to Wien's displacement theorem
involving the existence of one unknown thermodynamic potential function
$\phi(\omega/T)$ depending upon the oscillator frequency $\omega$ divided by
the thermodynamic temperature $T.$ \ At that time it was noted that the
unknown function $\phi(\omega/T)$ allowed two natural limits which made the
oscillator energy $U(\omega,T)$ independent of one of its two variables; the
limits corresponded to energy equipartition $U\rightarrow k_{B}T$ at high
temperature, and zero-point energy $U\rightarrow\hbar\omega/2$ at low
temperature. \ It was then suggested that \textquotedblleft the smoothest
interpolation\textquotedblright\ between these two limits led to the Planck
spectrum. \ Although the analysis may be compelling physics, the idea of a
\textquotedblleft smoothest interpolation\textquotedblright\ is not standard mathematics.

In the present article, we have attempted to go back to the question of the
blackbody spectrum within classical physics from the vantage point of all
three laws of thermodynamics, now including the third law which was not used
by Wien or Planck at the turn of the 20th century. \ We have attempted to use
simply the laws of thermodynamics and mathematical function theory to obtain
as much information as possible. We find that the third law of thermodynamics
forces the idea of zero-point energy for the classical harmonic oscillator.
\ Of course, the idea of a zero-point energy is now associated in the minds of
many physicists with quantum theory. \ However, our work does not require any
assumption regarding discrete energy processes. \ In addition to requiring
that an oscillator have a non-zero energy in thermal equilibrium at
thermodynamic absolute zero, the laws of thermodynamics provide sufficiently
stringent conditions that one can use them alone to derive the Planck
blackbody radiation spectrum including zero-point radiation. \ Finally, we
point out that the inclusion of classical zero-point energy, which is
automatic from the thermodynamic point of view, involves more natural force
ideas than the Planck expression which omits the zero-point energy. \

(revised September 10, 2018)


\begin{thebibliography}{99}
\bibitem {therm}\ I have yet to find a single textbook which considers the
thermodynamics of the harmonic oscillator aside from its appearance in the
application of statistical mechanics.
\bibitem {SS150}Our equation (\ref{SS69}) is analogous to Eq. (6-9) on page
150 of Sears and Salinger in ref. 9.
\end{thebibliography}
\end{document}